\documentclass[12pt]{article}
\usepackage{graphicx}

\newcommand\pubnumber{}
\newcommand\pubdate{\today}

\def\nara{Physics Department\\
Nara Women's University, \\
Kitauoya Nishi-machi, Nara, Japan -630 8506.}

\def\Journal#1#2#3#4{{#1} {\bf #2}, #3 (#4)}  
\def\Title#1{\begin{center} {\Large #1 } \end{center}}
\def\Author#1{\begin{center}{ \sc #1} \end{center}}
\def\Address#1{\begin{center}{ \it #1} \end{center}}

\newcommand\pubblock{\rightline{\begin{tabular}{l} \pubnumber\\
         \pubdate  \end{tabular}}}
\newenvironment{Abstract}{\begin{quotation}  }{\end{quotation}}
\newenvironment{Presented}{\begin{quotation} \begin{center} 
             PRESENTED AT\end{center}\bigskip 
      \begin{center}\begin{large}}{\end{large}\end{center} \end{quotation}}
\def\Acknowledgements{\bigskip  \bigskip \begin{center} \begin{large}
             \bf ACKNOWLEDGEMENTS \end{large}\end{center}}

\def\PLB{{\em Phys. Lett.}  B}
\def\PRL{\em Phys. Rev. Lett.}
\def\PRD{{\em Phys. Rev.} D}

\def\PR{{\em Phys. Rep.}}

\def\beq{\begin{equation}}
\def\eeq#1{\label{#1}\end{equation}}
\def\eeqn{\end{equation}}

\def\beqa{\begin{eqnarray}}
\def\eeqa#1{\label{#1}\end{eqnarray}}
\def\eeqan{\end{eqnarray}}

\let\bar=\overbar

\def\Dslash{\not{\hbox{\kern-4pt $D$}}}
\def\dslash{\not{\hbox{\kern-2pt $\del$}}}

\def\msb{{\bar{\ssstyle M \kern -1pt S}}}

\begin{document}
\begin{titlepage}
\pubblock

\vfill
\Title{Search for a $J/\psi\eta$ resonance \\in $B^{\pm}\to J/\psi\eta K^{\pm}$ decays at Belle}
\vfill
\Author{Tomoko Iwashita (for the Belle Collaboration)}
\Address{\nara}
\vfill
\begin{Abstract}
We report study of $B^{\pm}\to J/\psi\eta K^{\pm}$ decays at Belle.
In this analysis we search for $X(3872)$ as well as other narrow resonances in the $J/\psi\eta$ final state.
\end{Abstract}
\vfill
\begin{Presented}
The $5^{\rm th}$ International Workshop on Charm Physics\\
(Charm 2012)\\
14-17 May 2012, Honolulu, Hawai'i 96822.

\end{Presented}
\vfill
\end{titlepage}
\def\thefootnote{\fnsymbol{footnote}}
\setcounter{footnote}{0}

\section{Introduction}
\quad $X(3872)$ was first discovered in $J/\psi\pi^+\pi^-$ decay at Belle \cite{disx3872}.
It has been confirmed by CDF \cite{cdf}, D0 \cite{d0}, BaBar \cite{babar}, LHCb \cite{lhcb} and CMS \cite{cms} experiments.
Since $X(3872)$ mass ($M=3871.1\pm0.2$ MeV) is near the $D{\bar D^*}$ threshold and its narrow width ($\Gamma<1.2$ MeV) makes it a good candidate for $D{\bar D^*}$ molecule \cite{abx3872}.
A few other models such as tetra-quark model \cite{tetra}, $c{\bar c}g$ hybrid meson \cite{hybrid}, and vector glueball models are also suggested \cite{glball}.
Recent search for the charged tetra-quark partner in the $J/\psi\pi^+\pi^0$ final state resulted in a negative confirmation \cite{recentx3872}.
But still it is hard to totally rule out tetra-quark interpretation for $X(3872)$, as some model predicts $X(3872)^+$ to be broad thus it is difficult to observe with the currently available statistics. \\
\quad On the other hand, in either molecule or tetra-quark pictures, $X(3872)$  can have a $C$-odd partner which may dominantly decay into $J/\psi\eta$ final state.
A previous search was carried by BaBar, where they observe $B^{\pm}\to J/\psi\eta K^{\pm}$ decay mode but didn't find any signal for $X(3872)\to J/\psi\eta$ decay using the data corresponding $90\times10^{6}$ $B{\bar B}$ \cite{babarjek}.
With the Belle data corresponding to 8 times more statistics, we can either observe $C$-odd partner of $X(3872)$ or provide much tighter constraint.

\section{Reconstruction}
\quad $B$ meson is reconstructed using $B^{\pm}\to J/\psi\eta K^{\pm}$ decay mode.
The results presented here are obtained from the data sample corresponding to $772\times10^6$ $B{\bar B}$ events collected by the Belle detector \cite{belledet} at the KEKB energy asymmetric $e^+e^-$ collider \cite{kekb} operated at the $\Upsilon(4S)$ resonance. \\
\quad $J/\psi$ meson is reconstructed in its decay to $\ell^+\ell^-$ ($\ell$ $=$ $e$ or $\mu$).
Among the reconstructed charged particles, $e^{\pm}$ are mainly identified by the ratio between energy detected by the electromagnetic calorimeter (ECL) and momenta measured by tracking devices.
The $\mu^{\pm}$ candidates are identified by the hits recorded in the RPC layers interleaved in the iron flux return (KLM).
The photons are reconstructed from the energy deposits in electromagnetic calorimeter by requiring no matching charged track exists. 
In $e^+e^-$ decays, the four-momenta of all photons within 50 mrad of each of the original $e^+$ or $e^-$ tracks are included in the invariant mass calculation, in order to reduce the radiative tail. 
The reconstructed invariant mass of the $J/\psi$ candidates is required to satisfy 2.95 GeV$/c^2$ $<$ $M_{e^+ e^-(\gamma)}$ $<$ 3.13 GeV$/c^2$ or 3.04 GeV$/c^2$ $<$ $M_{\mu^+ \mu^-}$ $<$ 3.13 GeV$/c^2$.  
A mass- and vertex-constrained fit is performed to all the selected $J/\psi$ candidates in order to improve the momentum resolution. 
The $\eta$ candidates are reconstructed by combining two photons and we require 510 MeV$/c^2$ $<$ $M_{\gamma\gamma}$ $<$ 575 MeV$/c^2$.  
To reduce the background from $\pi^0 \to \gamma \gamma$, we reject the photon which in combination with another photon, in that event, gives mass in the region around $\pi^0$ mass defined as 117 MeV$/c^2$ $<$ $M_{\gamma\gamma}$ $<$ 153 MeV$/c^2$.  
Again mass-constrained fit is performed to all the selected $\eta$ candidates in order to improve the momentum resolution. 
Charged kaons are identified using momentum measurement as well as specific ionizations in the central drift chamber (CDC), time-of-flight (TOF) and aerogel Cherenkov counters (ACC). 
To reconstruct the $B$ candidates, we combine $J/\psi$, $\eta$ and kaon candidates.
To identify the $B$ candidate, two kinematic variables are used : energy difference $\Delta E \equiv E^*_B-E^*_{\rm beam}$ and beam-energy constrained mass $M_{\rm bc} \equiv \sqrt{(E^*_{\rm beam})^2-(P^{\rm cms}_B)^2}$.
Where, $E^*_{\rm beam}$ is the center-of-mass frame (cms) beam energy, and $E^*_B$ and $P^*_B$ are the cms energy and momentum of the reconstructed $B$ candidates.
In case of multiple candidates, the best one is selected by the $\chi^2$ based on the reconstructed masses of $J/\psi$ and $\eta$ candidates as well as kaon identification information.\\
\quad To suppress continuum background, events having a ratio of the second to zeroth
Fox-Wolfram moments~\cite{r2} $R_2 > 0.5$ are rejected.
Large $B\to J/\psi X$ MC samples, which is corresponding to $100$ times the data sample are used to understand the background. 
The non-$J/\psi$  background is estimated by the $M_{\ell\ell}$  sideband events in data. 
$B^{\pm} \to J/\psi \eta K^{\pm}$ yields are extracted from a 1D unbinned maximum likelihood fit performed to the $\Delta E$ distribution. 
For identification of resonance, a fit to $M_{J/\psi\eta}$ distribution is performed.

\section{Result}
\begin{figure}[h!]
\begin{center}
\includegraphics[angle=0,width=0.6\textwidth]{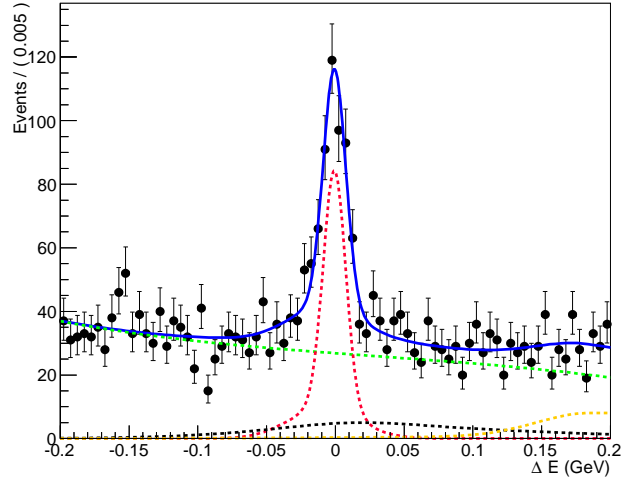} 
\caption{$\Delta E$ distribution of $B^{\pm}\to J/\psi\eta K^{\pm}$ candidates in 5.27 GeV$/c^2$ $<$ $M_{\rm bc}$ $<$ 5.29 GeV$/c^2$.}
\label{deltae}
\end{center}
\end{figure}

\begin{figure}[h!]
\begin{center}
\includegraphics[angle=0,width=0.6\textwidth]{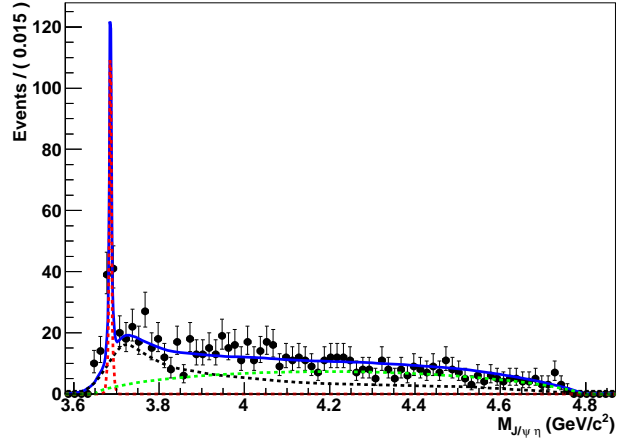} 
\caption{$J/\psi \eta$ mass distribution for the $B^{\pm}\to J/\psi\eta K^{\pm}$ signal candidates. The curves shows the signal (red for $B^{\pm}\to\psi'K^{\pm}$ and green for other $B^{\pm}\to J/\psi\eta K^{\pm}$) and the background component (black).}
\label{jpsieta}
\end{center}
\end{figure}

\begin{figure}[h!]
\begin{center}
\includegraphics[angle=0,width=0.6\textwidth]{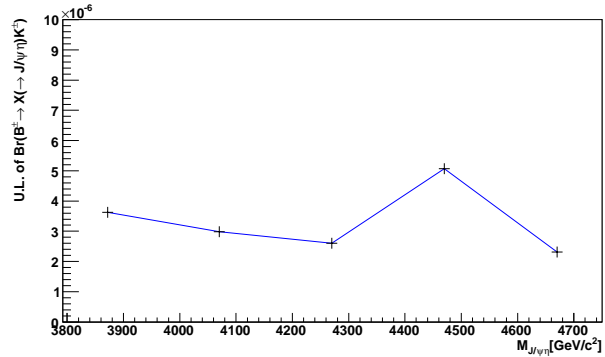} 
\caption{U.L. (@ 90\% C.L.) on the ${\cal B}(B^{\pm}\to X K^{\pm})\cdot {\cal B}(X\to J/\psi\eta)$ estimated at different masses using narrow width hypothesis.}
\label{ul}
\end{center}
\end{figure}

\quad $B^{\pm}\to J/\psi\eta K^{\pm}$ signal extraction by the fit performed to $\Delta E$ distribution as shown in Figure \ref{deltae}. 
The candidate events found in the range of -35 MeV $<$ $\Delta E$ $<$ 30 MeV are selected to identify resonances in $M_{J/\psi\eta}$ distribution. 
We observe a clear peak of $\psi'\to J/\psi\eta$ while there is no other narrow resonance as shown in Figure \ref{jpsieta}. 
Except for $\psi'\to J/\psi\eta$ contribution, the $B$ decay signal distribution is mostly explained by the three-body phase space.
We checked the branching fraction for $B^{\pm} \to \psi'K^{\pm}$ in $\psi'\to J/\psi\eta$ mode, and found that it is consistent with PDG \cite{pdg}. 
Excluding $B^{\pm}\to\psi'K^{\pm}$, $B^{\pm} \to J/\psi\eta K^{\pm}$ branching fraction is obtained to be ($1.2\pm0.1({\rm stat})\pm0.1({\rm syst})$). 
Since we didn't find any signal for $X(3872) \to J/\psi \eta$, we provided 90\% confidence level (C.L.) upper limit (U.L.) to be ${\cal B}(B^\pm \to X(3872) K^{\pm})\cdot{\cal B}(X(3872)\to J/\psi\eta)<3.8\times10^{-6}$ using a frequentist approach. 
Table \ref{table} summarizes the results of branching fraction measurement.
We also search for narrow resonance at different $M_{J/\psi \eta}$ points and provide U.L. (@ 90\% C.L.) as shown in Figure \ref{ul}.

\begin{table}[!h]
\begin{center}
\begin{tabular}{|l|c|c|}
\hline
Decay mode&Yield&${\cal B}$($\times10^{-4}$)\\
\hline
$B^{\pm}\to \psi' K^{\pm}$ (in $\psi'\to J/\psi\eta$)&52.0$\pm$8.2&5.8$\pm$0.9$\pm$0.4\\
$B^{\pm}\to J/\psi\eta K^{\pm}$ (excluding $\psi'K^{\pm}$)&395.0$\pm$26.0&1.2$\pm$0.1$\pm$0.1\\
\hline
\end{tabular}
\caption{Signal yields for decay mode within window of -35 MeV $<$ $\Delta E$ $<$ 30 MeV and branching fraction (${\cal B}$) of each decay process. First and second errors are statistical and systematic uncertainties respectively.}
\label{table}
\end{center}
\end{table}

\section{Summary}
\quad We observed $B^{\pm}\to J/\psi\eta K^{\pm}$ decay using $771\times10^{6}$ $B \bar B$ pair.
${\cal B}$$(B^{\pm}\to J/\psi\eta K^{\pm})$ (excluding $B^{^\pm}\rightarrow\psi' K^{\pm}$) is obtained to be $(1.2\pm0.1({\rm stat})\pm0.1({\rm syst}))\times10^{-4}$.
In our search for $C$-odd partners of $X(3872)$, we don't find any significant signature of narrow resonance and provided much tighter constraint than before on the U.L. as ${\cal B}(B^{\pm}\to X(3872)K^{\pm})\cdot{\cal B}(X(3872)\to J/\psi\eta)<3.8\times10^{-6}$ at 90$\%$ C.L.

\Acknowledgements
Authors's participation to the 5th International Workshop on Charm Physics (Charm2012) was surpported by NEXT KAKENHI, Grant-in-Aid for Scientific Research on Innovative Areas, entitled Elucidation of New hadrons with a Variety of Flavors.

\end{document}